\documentclass[prl,unsortedaddress,superscriptaddress,showpacs,twocolumn,floatfix]{revtex4}
\usepackage{amsmath}

\usepackage[dvips]{pstcol}
\usepackage{graphicx}
\newcommand{\lpc}{LPC-Caen, ENSICAEN, Universit\'e de Caen Basse-Normandie,
CNRS/IN2P3-ENSI, Caen, France}
\newcommand{\kul}{Instituut voor Kern- en Stralingsfysica, Katholieke
Universiteit Leuven, B-3001 Leuven, Belgium}

\begin{document}
\title{Determination of $|V_{ud}|$ from nuclear mirror transitions}

\affiliation{\lpc}
\affiliation{\kul}

\author{O.~Naviliat-Cuncic}\affiliation{\lpc}
\author{N.~Severijns}\affiliation{\kul}

\date{\today}

\begin{abstract}
The $V_{ud}$ element of the Cabibbo-Kobayashi-Maskawa quark mixing
matrix has traditionally been determined from the analysis of data
in nuclear superallowed $0^+\rightarrow 0^+$
transitions, neutron decay and pion beta
decay. We show here that this element can independently be
determined from nuclear mirror transitions. The extracted value,
$|V_{ud}| = 0.9719\pm 0.0017$, is at 1.2 combined standard deviations
from the value obtained in superallowed $0^+\rightarrow 0^+$
transitions and has a similar precision
than the value obtained from neutron decay experiments. We discuss
some prospects to improve its precision through experiments in
nuclear mirror transitions.
\end{abstract}


\maketitle
%
%
The unitarity conditions of the Cabibbo-Kobayashi-Maskawa (CKM)
quark mixing matrix \cite{cabibbo63,kobayashi73} provide sensitive
means to test the consistency of the three generation standard
electroweak model and to search for new physics beyond. A
stringent test is obtained from the elements of the first row
\begin{equation}
V_{ud}^2 + V_{us}^2 + V_{ub}^2 = 1
\label{eq:unitarity}
\end{equation}
where $V_{uj}$ denotes the amplitude of the quark mass eigenstate $j$ into
the quark weak eigenstate $d'$. The accuracy
in the verification of this
condition is largely due to the dominant value and
error of the $V_{ud}$ element which is obtained from weak decay
processes involving the lightest quarks.

Three traditional sources to determine $|V_{ud}|$ from experiments
have been considered during the past decades, namely, nuclear
superallowed $0^+\rightarrow 0^+$
pure Fermi transitions, neutron decay and pion beta
decay, and these have regularly been reviewed
\cite{Towner98,Towner03,Hardy07}.

The partial half-lives of nine nuclear superallowed $0^+\rightarrow 0^+$ transitions
have been studied in great detail \cite{hardy05,towner08}. Measurements
of lifetimes, masses and branching ratios have reached precisions
such that the required inputs for the calculation of the ${\cal F}t$
values were obtained at a
level of few parts in $10^{-4}$, yielding the value \cite{towner08}
\begin{equation}
|V_{ud}| = 0.97418(26)  \  {\rm (superallowed \ 0^+\rightarrow
0^+)} , \label{eq:Vud_superallowed}
\end{equation}
The value of $|V_{ud}|$ deduced from $0^+\rightarrow 0^+$ transitions is
since many years \cite{Towner98} dominated by uncertainties in
theoretical corrections. Present experimental activities are
oriented toward further reducing these uncertainties by testing
the calculations in other transitions \cite{towner08}.

Neutron decay involves both the vector and the axial-vector
interactions so that the determination of $|V_{ud}|$ from neutron
decay data, although free of nuclear structure corrections,
requires the analysis of at least two observables. The most
precise determinations have so far been obtained by combining the
neutron lifetime with the beta asymmetry parameter. The first
determination of $|V_{ud}|$ using only neutron decay data
\cite{Thompson90} yielded the value $|V_{ud}| = 0.9790(30)$. The
present world average recommended value for the neutron lifetime,
$\tau_n$ = 885.7(8) s \cite{amsler08}, combined with the world
average value for the beta asymmetry parameter, $A_n =
-0.1173(13)$ \cite{amsler08}, yields
\begin{equation}
|V_{ud}| = 0.9746(19) \  {\rm (neutron \ decay)} .
\label{eq:Vud_neutron}
\end{equation}
The improvement by a factor of about 1.5 over almost two
decades shows the difficulty of the associated experiments (see
e.g. \cite{Nico05,abele08}). Other results have however been
reported \cite{abele08,abele02} by taking selected values of the
most precise experimental data. Many experimental projects are
under way \cite{PPSN08} to improve the uncertainties on the
neutron lifetime and on several of the correlation parameters.

Finally, the absolute pion beta decay rate provides a clean
observable for the determination of $|V_{ud}|$. The main
experimental difficulty arises here from the very
weak ($10^{-8}$) branching of the beta decay channel. The most
recent experimental determination yields \cite{pocanic04}
\begin{equation}
|V_{ud}| = 0.9728(30) \ {\rm (pion \ decay)} \label{eq:Vud_pion} ,
\end{equation}
what is less precise than the value from neutron decay.

We consider here a new source to determine
$|V_{ud}|$, namely, the beta decay transitions between
$T=1/2$ isospin doublets in mirror nuclei. Such transitions are
sometimes called ``mirror decays'' and similarly to neutron decay
--which is the simplest mirror transition-- proceed via the vector
and axial-vector interactions. The principle to extract $|V_{ud}|$
from such transitions is then similar to that used in the analysis
of neutron decay, except for
the corrections associated with the nuclear system. The
corrections for the determination of the
${\cal F}t$ values in mirror transitions
have recently been surveyed \cite{Severijns08} and were
obtained with sufficient precision for their consideration in
the analysis reported here.
We use then below the results of this new survey and adopt
the definitions and notations given there, unless possible ambiguities
require it otherwise.

The vector part of the corrected statistical decay
rate function is given by \cite{Severijns08}
\begin{equation}
{\cal F}t \equiv f_V t(1 + \delta^\prime_R)(1 + \delta^V_{NS} - \delta^V_C)
\label{eq:Ft1}
\end{equation}
where $f_V$ is the uncorrected statistical rate function,
$\delta^\prime_R$ denotes nuclear dependent radiative corrections obtained
from QED calculations, $\delta^V_{NS}$ are nuclear structure
corrections and $\delta^V_C$ are isospin symmetry breaking
corrections for the vector contribution. For mixed Fermi/Gamow-Teller
transitions, ${\cal F}t$ is related to $V_{ud}$
by \cite{Severijns08}
\begin{equation}
{\cal F}t = \frac{K}{G^2_F V^2_{ud}} \frac{1}{C^2_V |M_F^0|^2
(1+\Delta^V_R)(1+f_A\rho^2/f_V)}
\label{eq:Ft2}
\end{equation}
where $K/(\hbar c)^6  =  2 \pi^3 ~\ln 2 ~ \hbar / (m_ec^2)^5$ and has the
value $K/(\hbar c)^6 = 8120.278(4) \times 10^{-10}$~GeV$^{-4}$s,
$G_F/(\hbar c)^3 = 1.16637(1) \times 10^{-5}$~GeV$^{-2}$ is the Fermi
constant \cite{amsler08},
$C_V = 1$ is the vector coupling constant,
$\Delta_R^V$ is a transition-independent radiative correction
\cite{marciano06}, $f_A$ is the statistical rate function for the axial-vector
part of the interaction, and $\rho$ is the Gamow-Teller to Fermi
mixing ratio. This ratio is defined by \cite{Severijns08}
\begin{eqnarray}
\rho & = & \frac{C_A M_{GT}^0}{C_V M_F^0} \left[ \frac{(1 +
\delta_{NS}^A - \delta_C^A)(1 + \Delta_R^A)} {(1 + \delta_{NS}^V -
\delta_C^V)(1 + \Delta_R^V)} \right]^{1/2} \nonumber \\
     & \approx & \frac{C_A M_{GT}^0}{C_V M_F^0} .
\label{eq:rho}
\end{eqnarray}
where the square root contains the nuclear structure, isospin
symmetry breaking and radiative corrections for the vector and
axial-vector contributions, $C_A$ is the axial-vector coupling
constant ($C_A/C_V \approx -1.27$) and $M_F^0$ and $M_{GT}^0$ are
the isospin symmetry limit values of the Fermi and Gamow-Teller
matrix elements, with $|M_F^0|^2$ = 1 for the $T_i = T_f = 1/2$
mirror transitions.

Using the corrected ${\cal F}t$ values from the recent
compilation \cite{Severijns08} it is possible to extract $|V_{ud}|$
from Eq.(\ref{eq:Ft2}) provided another observable, also
function of $\rho$, be known with sufficient precision. In the
present analysis we consider transitions where the
beta-neutrino angular correlation
coefficient, $a_{\beta\nu}$, and the beta decay asymmetry parameter,
$A_{\beta}$, have
been measured in the past. For $\beta^+$ mirror transitions,
their expressions as a function of the mixing ratio $\rho$,
in the limit of zero momentum transfer, are \cite{Severijns06}
\begin{equation}
a_{\beta\nu}(0) = \left( 1-\rho^2/3 \right) / \left( 1+\rho^2 \right)
\label{eq:a0}
\end{equation}
and
\begin{equation}
A_{\beta}(0) = \frac{\rho^2 - 2 \rho \sqrt{J(J+1)}}{(1+\rho^2)(J+1)}
\label{eq:A0}
\end{equation}
where $J$ denotes the spin of the initial and final states in the
transition.
At a precision level of about 1\%, as is the case
for the correlation coefficients we are dealing with here, the
impact of recoil
effects have however to be considered. To first order in recoil,
assuming the absence of second class currents
\cite{grenacs85} and time reversal invariance,
one then has for a $\beta^+$ transition
within a common isotopic multiplet \cite{holstein74}
\begin{equation}
a_{\beta\nu} = f_2(E)/f_1(E)  ,
\label{eq:a}
\end{equation}
and
\begin{equation}
A_{\beta} = f_4(E)/f_1(E) ,
\label{eq:A}
\end{equation}
with the spectral functions
\begin{eqnarray}
f_1(E) & = & a^2 + c^2 - \frac{2E_0}{3M} (c^2 - cb)  + \frac{2E}{3M}(3a^2 + \nonumber \\
       &   & + 5c^2-2cb) - \frac{2 m_e^2}{3E M} (c^2 -cb)  ,
\end{eqnarray}
\begin{eqnarray}
f_2(E) = a^2 - \frac{1}{3} c^2 + \frac{2 E_0}{3M} (c^2 - cb)
             -\frac{4E}{3M} (3c^2 - cb)  ,
\end{eqnarray}
and
\begin{eqnarray}
f_4(E) & = & \left( \frac{J}{J+1} \right)^{1/2}  \left[2ac - \frac{2 E_0}{3 M} (ac - ab) +
             \right. \nonumber \\
       &   & \left. + \frac{2 E}{3 M} (7ac - ab) \right] + \left( \frac{1}{J+1} \right)
             \left[c^2 + \frac{2 E_0}{3 M} (-c^2 +
             \right. \nonumber \\
       &   & \left. + cb) + \frac{E}{3 M} (-11c^2 + 5cb) \right] .
\end{eqnarray}
Here $E$ and $E_0$ denote respectively the total and the
total maximal positron energies, $M$ is the average mass of the
mother and daughter isotopes, and $m_e$ is the electron mass. In
this notation \cite{holstein74} $a$, $b$ and $c$ designate
respectively, the Fermi-, weak magnetism- and Gamow-Teller form
factors
\begin{equation}
a = g_V M_F , \ c = g_A M_{GT} .
\end{equation}
with $C_{i} = V_{ud} \ G_F \ g_{i}(q^2 \rightarrow 0)$,
($i = V,A$), $g_i$ being the vector and axial-vector form factors
and $q$ the momentum transfer. According to the
conserved-vector-current hypothesis \cite{holstein71,holstein74}
\begin{equation}
b = A \sqrt{(J+1)/J} M_F \ \mu .
\end{equation}
where $A$ is the mass number and
$\mu = [\mu(T_3) - \mu(T_3^\prime)]/ ( T_3 - T_3^\prime )$
%
%
is the isovector contribution to the magnetic moment, with $T_3$ the
third component of the isospin (in the convention where $T_3 = +1/2$
for a proton) and $\mu(T_3)$ and $\mu(T_3^\prime)$ the
magnetic moments of the mother and daughter nuclei.

The extraction of $|V_{ud}|$ proceeds then by solving Eqs.(\ref{eq:a}) or
(\ref{eq:A}) for $\rho$
and then inserting its value in Eq.(\ref{eq:Ft2}) with
the corresponding ${\cal F}t$ value from Ref.~\cite{Severijns08}
yielding
\begin{equation}
V_{ud}^2 = \frac{K^\prime}{{\cal F}t (1 + f_A \rho^2/f_V)} ,
\end{equation}
where $K^\prime = {K}/[G_F^2 \ C_V^2 \ ( 1 + \Delta_R^V )] =
5831.3(22)$~s,
and $\Delta_R = 2.361(38)$\% \cite{marciano06}. The error on $K^\prime$ is
dominated by the uncertainty on the radiative corrections $\Delta_R$.
The sign of $\rho$ was taken to be the same as in Ref.~\cite{Severijns08}.

The data included in the present analysis is summarized in Table~\ref{tab:data}.
The mirror transitions
considered here are those in $^{19}$Ne, $^{21}$Na and $^{35}$Ar, for
which the beta-neutrino angular correlation coefficient or the beta decay
asymmetry parameter have been measured.
The inclusion of recoil effects in the ${\cal F}t$ values was
found to have negligible effects on the resulting values for
$|V_{ud}|$. Corrections for $\delta_R^\prime$, $\delta_{NS}$ and
$\delta_C$ in the correlation coefficients cancel in the ratios
of the spectral functions, Eqs.(\ref{eq:a}) and (\ref{eq:A}).
Electromagnetic corrections \cite{holstein74}, other than the
dominant Coulomb effects contained in the
energy-dependent Fermi function $F(Z, E)$ and included in the
$f_{V,A}$ factors, were verified to be negligible at the present
level of precision.
The values for $E$ used in Eqs.(\ref{eq:a}) and (\ref{eq:A}) and
listed in Table~\ref{tab:data}, are average values
determined from the experimental conditions.

\begin{table}[!htb]
\begin{tabular}{lrrr}
\hline \hline
  & $^{19}$Ne~~~~ & $^{21}$Na~~~~ & $^{35}$Ar~~~~ \\
\hline
$a_{\beta\nu}$ & ---~~~~~~ & 0.5502(60)\footnotemark[1]  & ---~~~~~~ \\
$A_{\beta}$ & $-0.0391(14)$\footnotemark[2] & ---~~~~~~ & 0.430(22)\footnotemark[3] \\
${\cal F}t$ [s]\footnotemark[4] & 1718.4(32) & 4085(12) & 5688.6(72) \\
$f_A/f_V$\footnotemark[4] & 1.01428 & 1.01801 & 0.98938 \\
$E_0$ [MeV]\footnotemark[5] & 2.72783(30) & 3.03658(70) & 5.45514(70) \\
$E$ [MeV] & 0.510999 & 1.614(1) & 2.780(1) \\
$M$ [amu]\footnotemark[5] & 19.0001417(7) & 20.9957509(10) & 34.9720551(14) \\
$b$\footnotemark[6] &   $-148.5605(26)$   &   82.6366(27) & $-8.5704(90)$ \\
& & & \\
$\rho$  & 1.5995(46) & $-0.7136(72)$ & $-0.279(15)$ \\
$|V_{ud}|$ & 0.9716(22) & 0.9696(36) & 0.9755(38) \\
\hline \hline
\end{tabular}
\caption{Input data used to determine the values of $\rho$ and
$|V_{ud}|$ from the mirror transitions in $^{19}$Ne, $^{21}$Na and
$^{35}$Ar.}
\footnotetext[1]{From Ref.~\cite{Vetter08}.}
\footnotetext[2]{Value for E = m$_e$, from
Ref.~\cite{Calaprice75}.}
\footnotetext[3]{Weighted mean of values
from Refs.~\cite{Garnett88} and \cite{Converse93}.}
\footnotetext[4]{From Ref.~\cite{Severijns08}.}
\footnotetext[5]{Using data from Ref.~\cite{audi03}.}
\footnotetext[6]{Calculated with the magnetic moments listed
in Ref.~\cite{stone05}.}
\label{tab:data}
\end{table}

The beta asymmetry parameter in $^{19}$Ne decay has been measured
a couple of decades ago by the Princeton group
\cite{Calaprice75,Schreiber83}. Although the value reported in
\cite{Schreiber83} has a better precision than the results
quoted in \cite{Calaprice75}, we do not include here that input
since the result has never been published. From the value reported
in \cite{Calaprice75}, i.e. $A_{\beta} = -0.0391(14)$,
the value $\rho = 1.5995(46)$ is extracted, leading to
$|V_{ud}|(^{19}{\rm Ne}) = 0.9716(22)$.

A recent measurement of the beta-neutrino angular correlation
coefficient in $^{21}$Na decay produced the value $a_{\beta\nu} =
0.5502(60)$ \cite{Vetter08}. This result constitutes the most
precise measurement of this coefficient in a mirror transition.
The value of the mixing ratio extracted from this result is $\rho =
-0.7136(72)$ leading to $|V_{ud}|(^{21}{\rm Na}) = 0.9696(36)$.

Finally, in the decay of $^{35}$Ar, the beta asymmetry
parameter has reliably been measured twice, with the results $A_\beta =
0.49(10)$ \cite{Garnett88} and  $A_\beta = 0.427(23)$
\cite{Converse93}. The weighted mean of these
(Table \ref{tab:data}), which is dominated by the most precise of both,
yields a value $\rho = -0.279(15)$, leading to
$|V_{ud}|(^{35}{\rm Ar}) = 0.9755(38)$.

Except for $^{19}$Ne,
the recoil corrections appeared not to have a
significant impact in the determination of $\rho$.
For $^{19}$Ne,
Eq.~(\ref{eq:A0}) yields $\rho$ = 1.6015(44) what differs by
about half a standard deviation from the value quoted above.
For $^{21}$Na and $^{35}$Ar the values of $\rho$ obtained from
Eqs.~(\ref{eq:a0}) and (\ref{eq:A0}) are identical to those
given above.
The values of $\rho$ and $|V_{ud}|$ are also summarized in Table \ref{tab:data}.
The results obtained from $^{21}$Na
and $^{35}$Ar have comparable uncertainties,
which are a factor of 1.7
larger than the uncertainty on the value obtained from $^{19}$Ne.
The weighted mean of the three values is
\begin{equation}
|V_{ud}| = 0.9719(17)  \  {\rm (nuclear~mirror~transitions)}
\label{eq:Vud_mirrors}
\end{equation}
This result is consistent within 1.2 combined standard deviations with the
value obtained from nuclear superallowed $0^+\rightarrow 0^+$
transitions, Eq.(\ref{eq:Vud_superallowed}), and has an
uncertainty comparable to that obtained from neutron decay,
Eq.(\ref{eq:Vud_neutron}). This shows that nuclear mirror
transitions provide an independent sensitive source for the
determination of $|V_{ud}|$ and deserve therefore further
theoretical studies and experimental investigations to improve the
required inputs.

The survey in Ref.~\cite{Severijns08} reports, for each decay, the
contributions of the five inputs ($f_V$, lifetime, branching
ratio, $\delta_R$ and $\delta_C - \delta_{NS}$) to the error on
the ${\cal F}t$ values. For $^{19}$Ne the uncertainty is dominated
by experimental inputs to determine $f_V$ and by the lifetime. The
situation is similar in $^{21}$Na  where in addition, the
uncertainty on the branching ratio contributes at the third place.
Except for $^{35}$Ar, where all five inputs are known with a
relative uncertainty below $10^{-3}$, all other transitions
ranging from $^{3}$He to $^{45}$V have uncertainties dominated by
the experimental inputs. Improvements in the determination of
$|V_{ud}|$ offer therefore new opportunities for precision
experiments in mirror transitions. However, the uncertainties on
the values
of $|V_{ud}|$ given in Table~\ref{tab:data} are dominated
by those on $\rho$, so that improvements of these values call in
priority for new measurements of correlation coefficients.

We consider here in particular the impact of a new measurement of
$a_{\beta\nu}$ in $^{19}$Ne and $^{35}$Ar with a similar
precision to that achieved for $^{21}$Na
\cite{Vetter08}. For this purpose we use Eqs.(\ref{eq:a0}) and
(\ref{eq:A0}) instead of Eqs.(\ref{eq:a}) and (\ref{eq:A}).

\begin{figure}[!htb]
\begin{tabular}{cc}
\hspace*{-2mm}
\includegraphics[height=63mm]{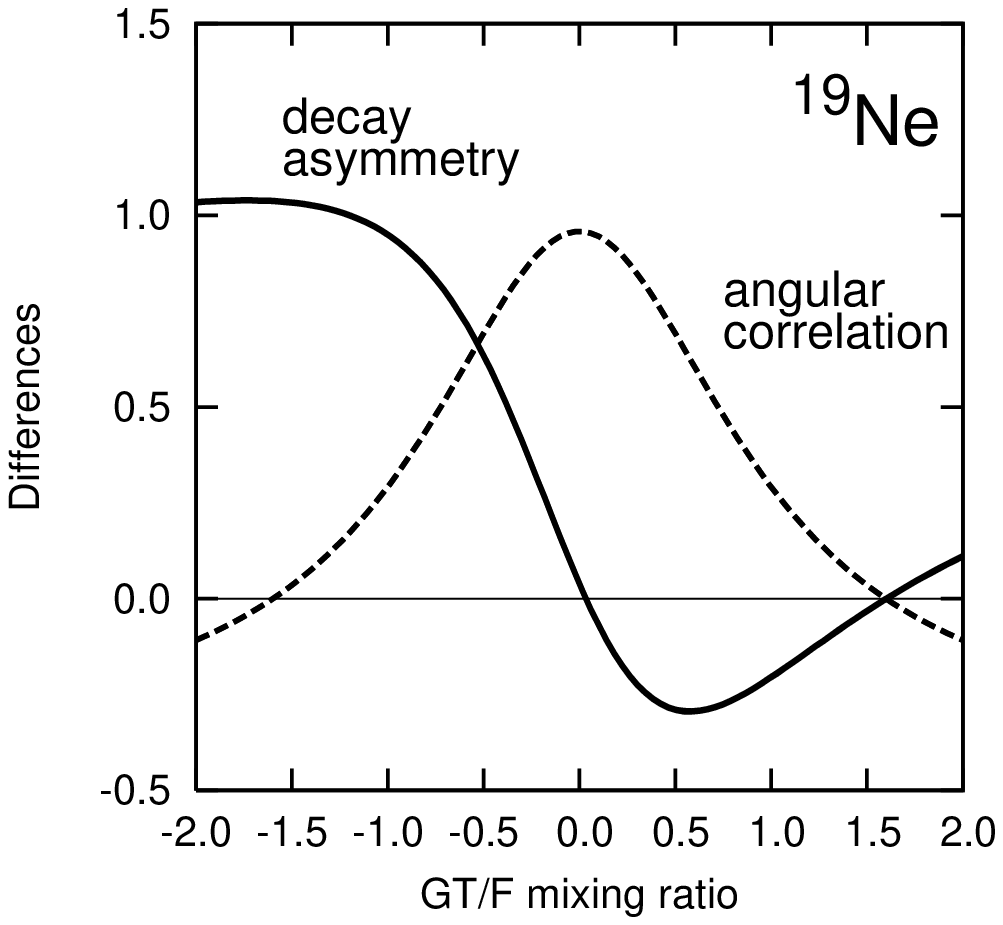}
&
\hspace*{-5mm}
\includegraphics[height=64mm]{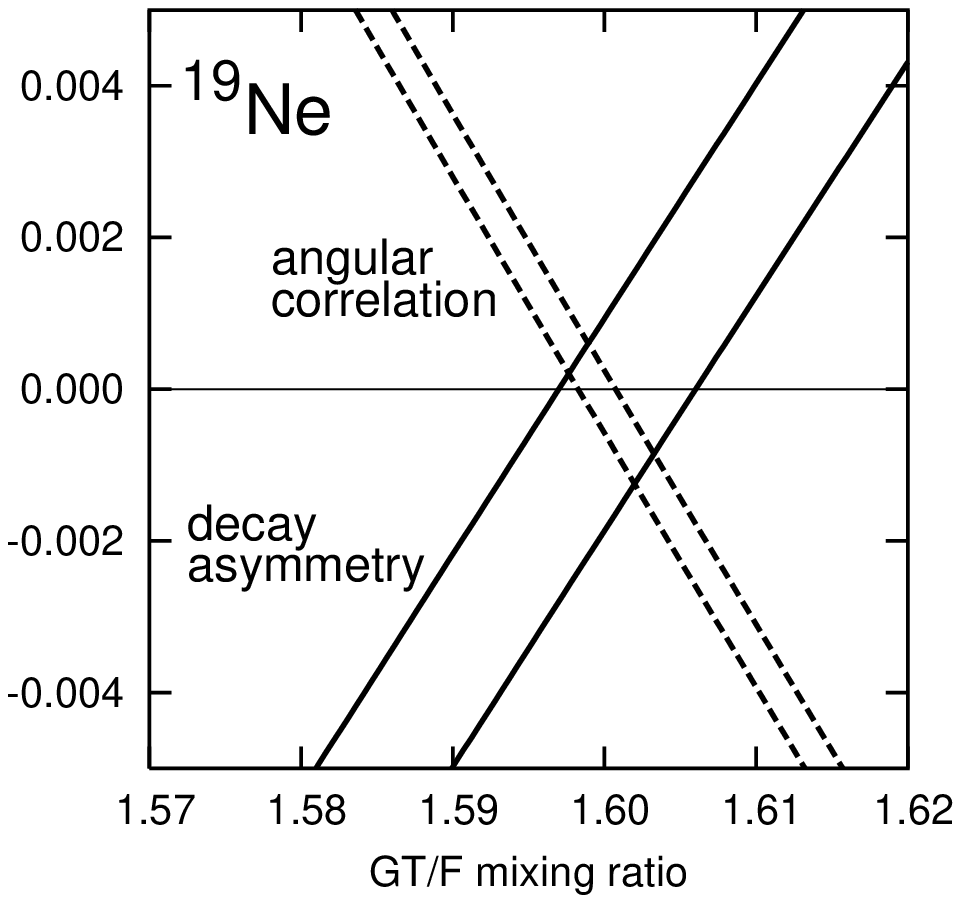}
\end{tabular}
\caption{Left panel: sensitivity of the angular correlation
coefficient and of the decay asymmetry parameter to the mixing ratio
in $^{19}$Ne decay. Right panel: intersection between
the two curves. The dotted lines indicate the region allowed by a
measurement of $a_{\beta\nu}$ with a relative uncertainty of
$\pm 1$\%.}
\label{fig:ne19}
\end{figure}

Figure \ref{fig:ne19} shows the sensitivity of
$a_{\beta\nu}$ and $A_\beta$ to $\rho$ for $^{19}$Ne decay. The solid
lines
are the differences between Eq.(\ref{eq:A0}) and the experimental
values (Table~\ref{tab:data}) at $\pm 1\sigma$. The dotted lines
are the differences between Eq.(\ref{eq:a0}) and the values of
$a_{\beta\nu}$ calculated with Eq.(\ref{eq:a0}) using the value of
$\rho$ given in Table~\ref{tab:data} and assuming a relative
precision of $\pm 1$\% on $a_{\beta\nu}$. The two solid and the two
dotted lines are superimposed on the left panel. The right panel
in Fig.\ref{fig:ne19} shows a zoom to the intersection region of
the two curves with zero. It is seen that a measurement of
$a_{\beta\nu}$ for $^{19}$Ne with a 1\% relative uncertainty
enables to reduce the uncertainty on $\rho$ by a factor of about
3. The sensitivities of the two observables to $\rho$ are
comparable.

\vspace*{-23mm}

\begin{figure}[!htb]
\begin{tabular}{cc}
\hspace*{-2mm}
\includegraphics[height=63mm]{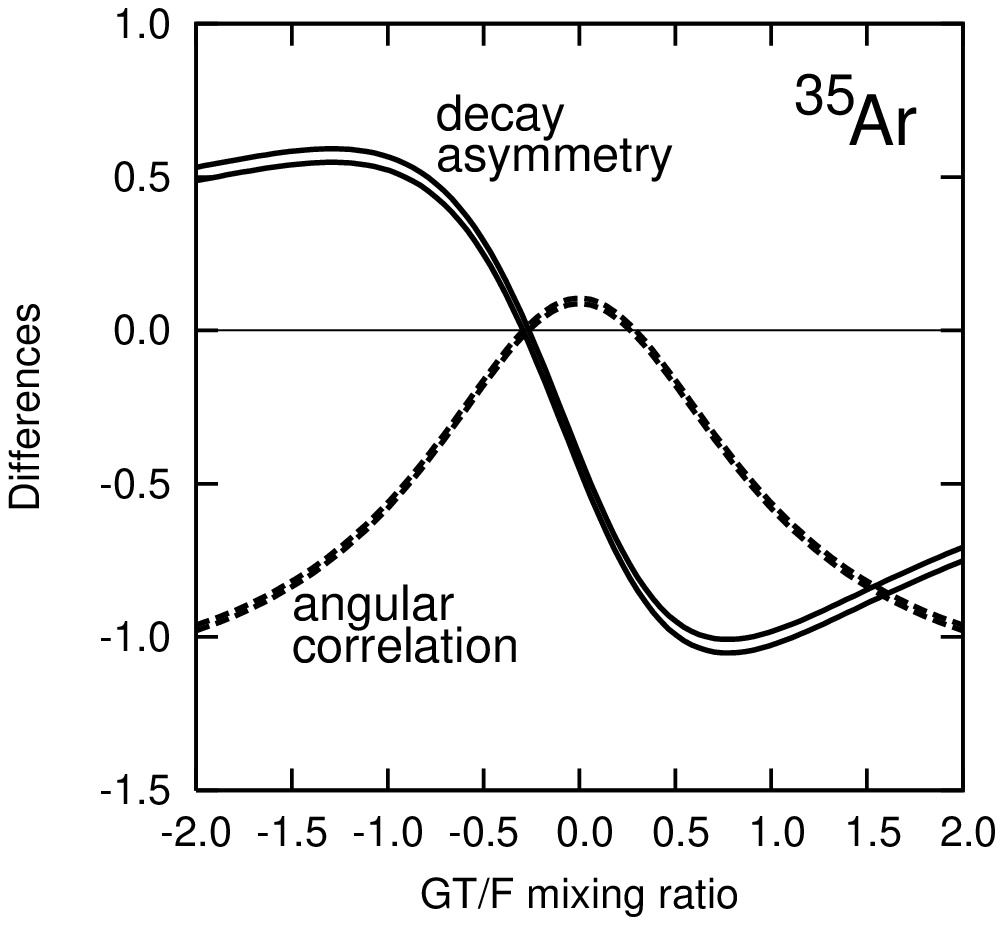}
&
\hspace*{-5mm}
\includegraphics[height=63mm]{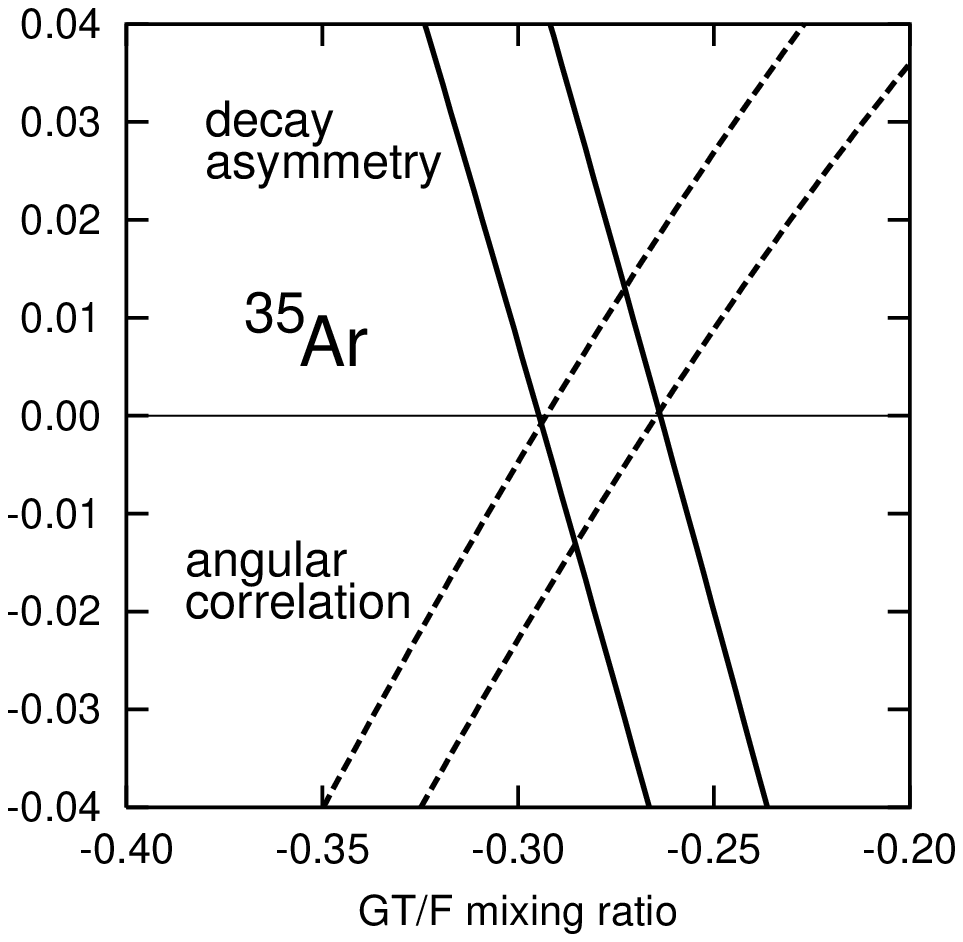}
\end{tabular}
\caption{Same than Fig.~\ref{fig:ne19} but for $^{35}$Ar decay.}
\label{fig:ar35}
\end{figure}

Figure \ref{fig:ar35} shows the same analysis for $^{35}$Ar decay.
The improvement is here very moderate since the intersection of
the two curves with zero occurs in the region where the
sensitivity of $A_{\beta}$ to $\rho$ is the largest. In fact,
among all mirror transitions considered in
Ref.~\cite{Severijns08}, it appears that $a_{\beta\nu}$
shows the largest sensitivity to $\rho$ in $^{19}$Ne decay and the
smallest sensitivity in $^{35}$Ar.

\vspace*{-23mm}

Based on the values of ${\cal F}t$ and $\rho$ listed in
Ref.~\cite{Severijns08} we found that measurements of
$a_{\beta\nu}$ provide better prospects than $A_\beta$
to improve on the value
of $|V_{ud}|$ from mirror transitions, the highest sensitivities
being obtained for $^{3}$He, $^{17}$F, $^{19}$Ne and $^{41}$Sc.
In the same context, measurements of $A_\beta$ look of interest
only in $^{19}$Ne decay, the sensitivity being then
similar to that of $a_{\beta\nu}$ in the same decay,
as shown above.

In conclusion, we have deduced the value of the CKM matrix element
$|V_{ud}| = 0.9719\pm 0.017$ using only data from
transitions in $^{19}$Ne, $^{21}$Na and $^{35}$Ar.
This demonstrates that nuclear mirror transitions provide an
independent sensitive source for the determination of
$|V_{ud}|$. Further theoretical studies as well as
precise determinations of the experimental inputs, and in particular of
the correlation coefficients, are desirable.


\end{document}